\documentclass[twocolumn,pra,showpacs,nofootinbib]{revtex4}
\usepackage[sort&compress]{natbib}
\usepackage{dcolumn,bm,graphicx,amsmath}
\begin{document}
\title{
Bose-Einstein condensates of polar molecules: anisotropic
interactions = anisotropic mass }


\author{  Andrei Derevianko}
 \email{andrei_AT_unr.edu} \homepage{http://wolfweb.unr.edu/homepage/andrei/tap.html}

\affiliation{Physics Department, University of  Nevada, Reno, Nevada
89557}

\date{\today}

\begin{abstract}
So far the theory of Bose-Einstein condensates (BEC) of polar molecules
was based on an ad hoc generalization of equations for spherical
atoms. Here I adopt a rigorous pseudo-potential approach to low-energy dipolar
interactions and derive a non-linear mean-field
Schr\"{o}dinger equation for a harmonically-trapped condensate.  I show that,
effectively, the dipolar interactions alter molecular mass.
The resulting effective mass is anisotropic: to the leading order the mass is
altered only for the motion along the polarizing field.
For a typical BEC of spin-polarized magnetically-interacting
alkali-metal atoms
the effective atomic mass is reduced by 10\% from it's bare value.
For a BEC of polar molecules the mass may be reduced
by a factor of a 1,000.
\end{abstract}



\maketitle The ongoing quest for a Bose-Einstein condensate  of
polar molecules\cite{specIssuePolMol04} is stimulated by a
remarkable richness of the quantum phenomena associated with their
large dipole-dipole interactions. Inherently anisotropic, the
dipolar interactions are crucial to quantum information
processing\cite{Dem02} and to realizing exotic states of matter
(e.g., topological\cite{MicBreZol06} and
checker-board\cite{GorSanLew02} phases) with ultracold molecules.

%

{\em Background---} The intricacies of the  many-body problem are
rooted in inter-particle interactions
 that lead to non-separable Hamiltonians.
Introducing pseudo-potentials, i.e., effective interactions that
are simpler than the original interactions, makes the problem more
tractable. In physics of ultracold atoms, all seemingly disparate
interactions can be well approximated by a contact pseudo-potential.
Its strength is determined by  $s$-wave scattering length $a_{ss}$,
which is computed by solving low-energy scattering  problem
with the original potential.
This contact interaction is central to arriving at the non-linear
Schr\"{o}dinger equation for the condensate wavefunction $\Psi\left(
\mathbf{r}\right)$. The hallmark Gross-Pitaevskii
equation~\cite{GPEabbreviated} (GPE) reads:
\begin{eqnarray}
\left(  -\frac{\hbar^{2}}{2M}\mathbf{\Delta}+U\left(\mathbf{r}\right)
+g'_{0}\left\vert \Psi\left(  \mathbf{r}\right)  \right\vert ^{2}\right)
\Psi\left(  \mathbf{r}\right)  = \mu_0 \Psi\left(  \mathbf{r}\right)  ,
\label{Eq:GPE}%
\end{eqnarray}
where $\mu_{0}$ is the chemical potential, and $U\left(\mathbf{r}\right)$
is the external confinement potential.
The non-linear term arises due to interparticle
interactions,
$g'_{0}  =4\pi \hbar^{2}/M \,   a_{ss} $. Namely this non-linearity
yields a wealth of non-trivial effects, such as solitons,
in physics of the condensates~\cite{PetSmi02}.

Why do the dipolar interactions require going beyond the conventional
approximation? Compared to the
$1/r_{12}^6$  dependence of the conventional isotropic van der Waals interactions on interparticle separations,
the dipolar interactions are both long-range, $1/r_{12}^3$, and anisotropic.
This seemingly innocuous  power-law variation
crucially modifies low-energy collision process that underlies the
pseudo-potential formalism.
Contributions of partial waves beyond the $s$-wave no longer ``freeze out'' and
the  scattering is characterized by an infinite number
of scattering lengths. Due to the anisotropy, molecules exert torques on each other and
various spherical waves (e.g., $s$ and $d$) become coupled. These couplings result
 in additional ``anisotropic'' scattering lengths~\cite{Der03}.

Following Yi and You\cite{YiYou00} (YY),
the rapidly growing literature on dipolar BECs,
see e.g.~\cite{YiYou00,SanShlZol00,ODeGioEbe04,StuGriKoc05etal,PedSan05,CooRezSim05,RonBorBoh06,KawSaiUed06}, is
based on
an effective interaction that is represented as a sum
of the contact pseudo-potential and the classical
dipole-dipole interaction.
This ad hoc approximation has a shortcoming of being valid only in a perturbative (Born)
regime.
Another, both aesthetic and practical shortcoming,
is that in the YY approximation the GPE becomes
a non-linear {\em integro-differential} equation\cite{YiYou00} that lacks the
appealing minimalism
of Eq.~(\ref{Eq:GPE}). By contrast, here,
 starting
from the rigorous quantum-mechanical description of the dipolar
collision process~\cite{Der03},
we attempt to overcome both shortcomings: (i)
the employed pseudo-potential
involves scattering parameters that may be tuned  all the way
through the resonances
and (ii) the dipolar GPE derived here has a simple mathematical structure.

{\em Dipolar pseudopotential---} In  a typical dipolar BEC setup, a
molecular gas forms a cloud in an external harmonic trapping
potential.
Orientation of molecular dipoles $D$ is fixed by applying a polarizing E-field
(otherwise, molecular rotations would average dipole moments to zero). Then
as $r_{12}\rightarrow \infty$, the molecular interactions acquire dipolar character,
\begin{equation}
 V(\mathbf{r_{12}}) \rightarrow
 \frac{D^2}{|r_{12}|^3} (3 \cos^2 \theta_{12} -1)\, . \label{Eq:Vdd}
\end{equation}
Here $\theta_{12}$ is the angle between collision axis $\mathbf{r}_{12}$ and the
polarizing field.
The collision process is also determined by the short-range part of the potential:
as molecules approach each other, the electronic clouds start to overlap, and the
interactions substantially depart from the dipolar form (\ref{Eq:Vdd}).
The YY approximation treats the short- and long-range
parts of the full interaction on separate footings; we will incorporate
both consistently.

We need to describe a quantum dipolar collision process at ultralow temperatures.
I assume that the polarized atoms or molecules follow a
unique potential surface.  For example,
recently attained BEC of highly-magnetic chromium
is comprised of spin-polarized atoms~\cite{GriWerHen05}.
The atoms are trapped in the lowest-energy
Zeeman sublevel; transitions to the
upper-energy levels are forbidden energetically.
To quantify the
scattering, one has to solve a multi-channel problem. The relevant scattering
parameters are the following limits of the K-matrix characterizing couplings
between $\ell$ and $\ell'$ partial waves~\cite{Der03},
\begin{equation}
a_{\ell m; \ell' m'} = - \lim_{k\rightarrow 0}
\mathcal{K}_{\ell m \rightarrow \ell' m'}/{k} \,,
\end{equation}
where $\hbar k$ is the relative momentum of the colliding pair. K-matrix essentially
governs the asymptotic form of the scattering wavefunction for large interparticle
separations.
Long-range, $1/r_{12}^3$, character of the dipolar interaction ensures that the above limits
are finite.
The quantities $a_{\ell m; \ell' m'}$ have a dimension of length and will be referred
to as scattering lengths.

For illustration, consider  scattering lengths
for a pure dipolar interaction (i.e., assuming the validity of Eq.~(\ref{Eq:Vdd})
for all $r_{12}$)
in the Born approximation. The formalism is described in Ref.~\cite{Der03} and it is based on
a system of coupled radial equations for individual partial waves.
We find that in the Born approximation
both the diagonal and off-diagonal scattering lengths fall off
as $\ell^{-2}$ with increasing $\ell$.  In the following I assume
that the dominant effects are due to $s-s$ and $s-d$ scattering lengths,
$a_{ss}$ and $a_{sd}$. The former is mainly determined by the short-range
part of the potential and to the latter by the dipolar coupling. In the Born
 approximation~\cite{Der03}
\begin{equation}
a_{sd}^\mathrm{Born} = - 1/ (6 \sqrt{5}) \, M
D^2 /\hbar^2 \, . \label{Eq:asdBorn}
\end{equation}
Values of $a_{sd}^\mathrm{Born}$ for  molecules of present experimental
interest are listed in Table~\ref{Tab:molpars}.
The off-diagonal scattering length is strongly suppressed (by
$\sim (1/137)^2$) for magnetically-interacting atoms: it is
 -0.01 nm for Cs and -0.2 nm for Cr~\cite{GriWerHen05}.

\begin{table}[h]
\begin{center}
\begin{tabular}[c]{lccc}
\hline\hline
& $D$, Debye & $a_{sd}^{\mathrm{Born}}$, nm \\
\hline
OH  $\left(  X\,^{2}\!\Pi_{3/2}\right)  $ & 1.65 & -52 \\
RbCs $\left( X \,^{1}\!\Sigma\right)  $ & 1.2 & -350 \\
KRb $\left(  X\,^{1}\!\Sigma\right)  $ & 0.59 & -48 \\
NH $\left(  X^{3}\Sigma^{-}\right)  $ & 1.38 & -32 \\
\hline\hline
\end{tabular}
\end{center}
\caption{
Anisotropic scattering length
$a_{sd}$ in the Born approximation for molecules
of current experimental interest\cite{specIssuePolMol04}.
 \label{Tab:molpars} }
\end{table}

{\em Pseudopotential ---} The low-energy pseudopotential for
anisotropic scattering was introduced
in Ref.~\cite{Der03}.
An earlier variational BEC study with this pseudopotential (albeit its
erroneous version) was carried out in Ref.~\cite{YiYou04}. Recently,
there was a study of the validity of the pseudopotential approach~\cite{KanBohBlu07}.
These authors found that the pseudopotential description remains accurate as long as
$a_{sd}$ is smaller than the characteristic length of the trapping potential.

For the goals of this paper we simplify the rigorous pseudopotential~\cite{Der03}. More details of the discussion presented below can be found in Ref.~\cite{Der08_simppseudo}.
We assume that the global BEC properties
can be described by well-behaved wavefunctions. In this
case we may operate in terms of the momentum-space representation. The matrix element
of the  pseudopotential between two plane waves $\langle
\mathbf{r}|\mathbf{k}\rangle=\left(  2\pi\right)  ^{-3/2}e^{i\mathbf{k}%
\cdot\mathbf{r}}$ is given by $\bar{v}\left(  \mathbf{k},\mathbf{k}^{\prime
}\right)  $  of  Ref.~\cite{Der03}. For our  case of the dipolar
interactions truncated at $s-s$ and $s-d$ couplings it reads
$
\bar{v}\left(  \mathbf{k},\mathbf{k}^{\prime}\right)    =\frac{1}{2\pi^{2}%
}\frac{\hbar^{2}}{M}\left(  a_{ss}-a_{sd}~\mathcal{F}\left(  \mathbf{k},\mathbf{k}%
^{\prime}\right)  \right)$ with
\[
\mathcal{F}\left(  \mathbf{k},\mathbf{k}^{\prime}\right)  = \sqrt{5}\left\{
P_{2}\left(  \cos\theta_{k}\right)  +\left(  k/k^{\prime} \right)
^{2}P_{2}\left(  \cos\theta_{k^{\prime}}\right)  \right\}  \, ,
\]
where $\theta_{k}$ and $\theta_{k^{\prime}}$ are angles between the
polarizing field  and $\mathbf{k}$ and
$\mathbf{k}^{\prime}$.

Now, under simplifying assumption of harmonic trapping, I transform the
momentum-space expression back into the coordinate space.  We write for a
matrix element of the pseudopotential (cf. Ref.~\cite{Omo77} for Rydberg atoms)
\begin{eqnarray}
\langle\psi|\hat{V}_\mathrm{ps}|\psi\rangle&=&{\left(  2\pi\right)^{-3}}
\int
d\mathbf{k}d\mathbf{k}^{\prime}d\mathbf{r}d\mathbf{r}^{\prime} \times
\nonumber \\
&& \psi^{\ast
}\left(  \mathbf{r}^{\prime}\right)  e^{i\mathbf{k}^{\prime}\cdot
\mathbf{r}^{\prime}}\bar{v}\left(  \mathbf{k}^{\prime},\mathbf{k}\right)
\psi\left(  \mathbf{r}\right)  e^{-i\mathbf{k}\cdot\mathbf{r}} \, .
\label{Eq:Meltransform}%
\end{eqnarray}
Only certain values of $|\mathbf{k}|$ and $|\mathbf{k}^{\prime}|$
contribute to this integral.
Experimentally,
the collisions occur in
the presence of harmonic trapping potential, $U\left(  \mathbf{r}\right)
=\frac{1}{2}M\left(  \omega_{x}^{2}x^{2}+\omega_{y}^{2}y^{2}+\omega_{z}%
^{2}z^{2}\right)$. For two harmonically-confined particles the center-of-mass
and relative motions decouple and the Hamiltonian for the relative motion
reads $H_{r}=p_{12}^{2}/\left(  2\mu\right)  +\left(  \frac{\mu}{M}\right)
U\left(  \left\vert \mathbf{r}_{12}\right\vert \right)  +V\left(
\mathbf{r}_{12}\right)$, where $V\left(  \mathbf{r}_{12}\right)  $ is the
full interaction potential between the particles and $\mu=M/2$. In the
stationary problem we solve the eigenvalue equation $H_{r}\psi\left(
\mathbf{r}_{12}\right)  =E_{r}\psi\left(  \mathbf{r}_{12}\right)  $, $E_{r}$
being the energy of the relative motion. Ref.~\cite{BluGre02}
presented a numerical
comparison of solutions of this equation with the full molecular
potential and its pseudopotential representation (for isotropic $s$-wave
scattering).
They  observed that
for
velocity-dependent potentials the relevant collision momentum $k_c$  is
$\hbar^{2}k_{c}^{2}/(2\mu)=E_{r}$. Indeed,
the collision process occurs at
$\left\vert \mathbf{r}_{12}\right\vert $ much smaller than the harmonic length.
In this region
$U\left(  \left\vert \mathbf{r}_{12}\right\vert \right)  \approx0$ and
the kinetic energy is  $E_{r}$. This leads to
$k_{c}^{2}=3/2~M \bar{\omega}/\hbar$,  with $\bar{\omega}=\sum_i
\omega_i/3$ being the average of the three trapping frequencies.

Returning to the evaluation of the integral~(\ref{Eq:Meltransform}),
we see that the relevant
contributions are accumulated at $\left\vert \mathbf{k}\right\vert =\left\vert
\mathbf{k}^{\prime}\right\vert =k_{c}$. Representing $\cos\theta_{k}=k_{z}/k,$
and using the properties of the Dirac $\delta$-function, we arrive at%
\begin{eqnarray}
\hat{V}_\mathrm{ps}\left(  \mathbf{r}_{12}\right) & \approx & 4\pi\frac{\hbar^{2}}{M}%
\delta\left(  \mathbf{r}_{12}\right) \times  \label{Eq:PS} \\
&& \left\{  a_{ss}+\sqrt{5}a_{sd}\left[
1+\frac{3}{2}\frac{1}{k_{c}^{2}}\left(  \overleftarrow{\partial_{z_{12}}^{2}%
}+\overrightarrow{\partial_{z_{12}}^{2}}\right)  \right]  \right\}  .
\nonumber%
\end{eqnarray}
This is a contact interaction that depends on  the traditional
$s$-wave scattering length $a_{ss}$ and ``anisotropic'' scattering
length $a_{sd}$. The latter is due to the dipolar coupling of $s$
and $d$ partial waves. Both scattering lengths are to be determined
from multi-channel scattering calculations with the full interaction
potential. One recognizes the conventional contact interaction $4\pi
\hbar^{2}/M \, \delta\left(  \mathbf{r}_{12}\right)  a_{ss}$ for the
$s$-wave scattering; the remaining part, with $a_{sd}$, is due to
the dipolar interactions.  In this term the $\overleftarrow
{\partial_{z_{12}}^{2}}\equiv\overleftarrow{\partial^{2}/\partial z_{12}^{2}}$
acts on the bra and $\overrightarrow{\partial_{z_{12}}^{2}}$ operates on the
ket. The
dipolar contribution to $\hat{V}_\mathrm{ps}$ breaks into two parts:
isotropic and anisotropic (derivative) terms. The isotropic dipolar
contribution merely renormalizes the traditional spherically-symmetric
pseudopotential. It is the derivative term that defines the novel
physics of the dipolar  BEC.
The anisotropy is imposed by the
polarizing field (the derivatives are taken along the field).
The long-range character of the dipolar interactions is manifested through
$k_c^2$ which characterizes the entire trapping potential.


The low-energy dipolar interactions can be controlled by external fields.
As in the case of Feshbach-resonance  tunability of $a_{ss}$,
the anisotropic  length $a_{sd}$ may exhibit resonances. There is an
evidence for such a field-dependent resonance in dipolar collisions
of E-field-polarized Rb atoms~\cite{DebYou01}. This opens an
intriguing possibility of {\em resonantly} controlling and enhancing dipolar
interactions in BECs.
The described resonant scattering mechanism  is to be distinguished from a recent
proposal~\cite{GioGorPfa02} on controlling the strength
of the dipolar interaction by modulating the dipoles by oscillating
fields.



{\em Mean-field approximation ---} With the derived  pseudo-potential, I proceed to analyzing
properties of the dipolar BEC.
In the mean-field approximation, all identical bosons occupy the
same single-particle wavefunction $\Psi(r)$.
Under this assumption,
 we arrive at the total energy of the
condensate ($\Psi$
is normalized to the total number of particles $N$)
\begin{eqnarray}
E\left[\Psi\right]   &=&\int d\mathbf{r}
\left( \frac{\hbar^{2}}{2M}
\left\vert
\nabla\Psi\right\vert ^{2}+U\left(  \mathbf{r}\right)  \left\vert
\Psi\right\vert ^{2}+
\right. \nonumber
\\
& + &\left.
 \frac{1}{2}g_{0}\left\vert \Psi\right\vert ^{4}
-g_{d}\left(  \frac{\partial}{\partial z}\left\vert \Psi\right\vert
^{2}\right)  ^{2}\right) \, , \label{Eq:EtotPsi}%
\end{eqnarray}
where the isotropic and anisotropic coupling  parameters are defined as
\begin{eqnarray}
g_{0}  &=&4\pi \hbar^{2}/M \, \left(  a_{ss}+\sqrt{5}\, a_{sd}\right),
\label{Eq:g0} \\
g_{d}  &=& 2\pi \hbar^{3} / (M^{2}\bar{\omega}) \,\sqrt{5} \, a_{sd} \,.
\label{Eq:gd}
\end{eqnarray}
The dipolar effects are governed by the anisotropic
length $a_{sd}$.  Dipolar
interactions modify the traditional GP term
$\frac{1}{2}g_{0}\left\vert \Psi\right\vert ^{4}$ and
appear in the newly-introduced derivative term,
$-g_{d} (\partial_z | \Psi|^{2})^{2}$.

\begin{figure}[h]
\begin{center}
\includegraphics[scale=0.75]{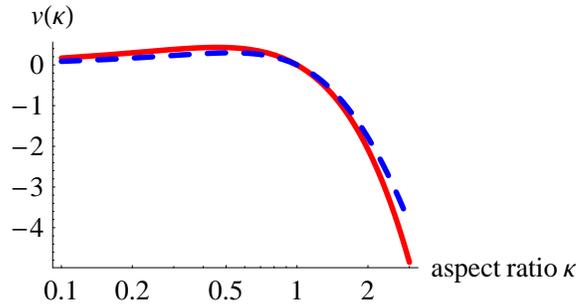}
\end{center}
\caption{Comparison of the dipolar interaction energy computed in
the traditional YY approximation (solid line)
and within the present pseudo-potential approach (dashed line).
The comparison is given as a function of the aspect ratio of the cloud.
 } \label{Fig:compare}
\end{figure}

In the present analysis, we truncated the expansion over
partial waves at the $s-s$ and $s-d$ scattering lengths.
To estimate the truncation error, we turn to the regime where the
Born approximation is valid. Then we use Eq.(\ref{Eq:asdBorn})
and compare the results with those computed within the conventional YY approximation.
In particular, consider an important expectation value $\langle V_{DD} \rangle$ of the
dipolar interaction
for an axially-symmetric ground state of the harmonic trapping potential
($\omega_x=\omega_y\equiv \omega_\perp$,
aspect ratio $\kappa=(\omega_z/\omega_\perp)^{1/2}$, polarizing field is along the axis of symmetry).
We may parameterize $\langle V_{DD} \rangle =  N \hbar \omega_\perp (a_{sd}^\mathrm{Born}/a_{\perp}) \times v(\kappa)
$.
In the YY approximation
$v^{YY}\left( \kappa \right) =
-6\kappa/(\sqrt{2}\pi) \, b\left( \kappa \right)$, where
function $b\left( \kappa \right)$ is given
in Ref.~\cite{YiYou00}. In the present pseudo-potential approach
$v^{ps}\left( \kappa \right) =\sqrt{\frac{5}{2\pi }}\left( \kappa -\frac{3\kappa ^{3}}
{(\kappa ^{2}+2)}\right) $. A comparison of the two reduced dipolar energies
$v^{ps}$ and $v^{YY}$ as functions of the aspect ratio
 is presented in Fig.\ref{Fig:compare}. We find a  good agreement between the
two curves.
Both energies vanish for spherically-symmetric ($\kappa=1$) traps.



{\em Effective anisotropic mass---} For real-valued $\Psi $, we  combine the kinetic-energy and the  derivative
terms in Eq.~(\ref{Eq:EtotPsi}). The result suggests introducing effective mass along the polarizing
field,
\begin{equation}
M_{zz}^\mathrm{eff}\left(  \mathbf{r}\right)= M/ (1- 8
g_{d} M ~n\left(  \mathbf{r}\right)/\hbar^{2} ) \, ,
\end{equation}
where the number density $n(r) = |\Psi(r)|^2$. The mass remains ``bare'' ($M$) for the
motion perpendicular to the polarizing field. In the Born approximation, the
relative change in the mass is
\[
   \frac{\delta M_{zz}^\mathrm{eff} }{M } \approx - \frac{8\pi}{3} \frac{n(r) D^2}{ \hbar \bar{\omega}} \, ,
\]
i.e., the mass is reduced from it's bare value by the ratio of the characteristic dipole interaction $n(r) D^2$ to
the trapping energy. As an illustration, consider a BEC of alkali-metal atoms
(magnetic moment $1 \, \mu_B$)
with $n \approx 10^{14} \,  1/\mathrm{cm}^3$ in a trap of
$\bar{\omega}=2\pi \times 100 \, \mathrm{Hz}$. We find that the  mass
is reduced by 10\%.
For a BEC of molecules the effect is even more pronounced. Here the effective
mass is about a $1,000$ times smaller than the bare mass.


\begin{figure}[h]
\begin{center}
\includegraphics[scale=0.75]{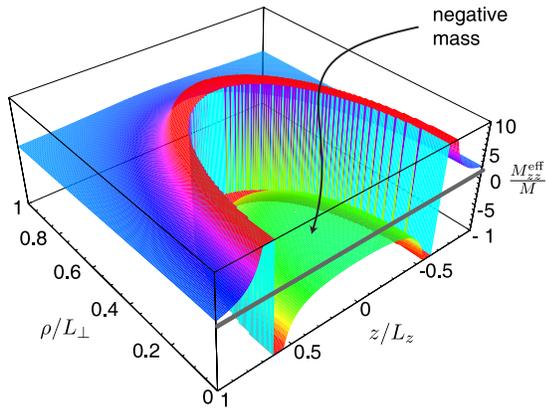}
\end{center}
\caption{ Effective anisotropic mass
$M_{zz}^\mathrm{eff}(\mathbf{r})$ as a function of position for an
``anomalous'' ($a_{sd} > 0$) BEC in a harmonic trapping potential.
 An unperperturbed Gaussian density profile of cylindrical symmetry
 about the polarizing filed
  is assumed,
 $n(\mathbf{r})= n_{\max} \exp\{- (\rho/L_\perp)^2  - (z/L_z)^2 \}$.
The peak density is choosen to exceed the critical value, $n_{\max}
> \hbar^2/(8 M g_d)$,
 and the
effective mass $M_{zz}^\mathrm{eff}$ is negative in the center of
the cloud. As we move to the outer regions of the cloud, $n$
decreases, $M_{zz}^\mathrm{eff}$ goes through a singularity and becomes
positive. For even smaller densities, $M_{zz}^\mathrm{eff}$
approaches the bare mass $M$.
 } \label{Fig:massCoral}
\end{figure}

The effective mass depends on the local density. Especially
interesting is the case when $a_{sd}$ is tuned to positive values:
here the effective mass may become {\em negative} (see Fig.~\ref{Fig:massCoral}).
Here the condensate is unstable in the region of
negative mass: the energy may be
lowered by accumulating density oscillations of increasingly shorter
wavelengths. By requiring that at the peak density
$M_{zz}^\mathrm{eff}>0$, we find that the BEC is stable for a number
of molecules below the critical number
$
N <  N_\mathrm{crit} \approx
1/8 {\sqrt{\pi/5}}  \sqrt{\hbar/(M \, \omega_z )} / a_{sd} .
$
Such an instability is not present in the traditionally-considered
dipolar gases, as $a_{sd}^\mathrm{Born} < 0$. We observe that
the instability in this case is likely to be related to the well-known
instability due to attractive
isotropic interactions~\cite{PetSmi02} (when the
effective scattering length $a_{ss} + \sqrt{5} a_{sd} < 0$).

Finally,  we minimize Eq.~(\ref{Eq:EtotPsi}) with respect to $\Psi$,
and arrive at the  non-linear Schr\"{o}dinger equation
\begin{eqnarray}
\lefteqn{
\left(  -\frac{\hbar^{2}}{2M}\mathbf{\Delta}+U\left(\mathbf{r}\right)
+g_{0}\left\vert \Psi\left(  \mathbf{r}\right)  \right\vert ^{2}\right)
\Psi\left(  \mathbf{r}\right) +} \nonumber  \\
 &+&g_{d}\left(  \frac{\partial^{2}}{\partial
z^{2}}\left\vert \Psi\left(  \mathbf{r}\right)  \right\vert ^{2}\right)
\Psi\left(  \mathbf{r}\right)  = \mu_{0}\Psi\left(  \mathbf{r}\right)  .
\label{Eq:DGPE}%
\end{eqnarray}
This equation subsumes the traditional GPE~(\ref{Eq:GPE}) when $a_{sd}=0$, i.e.,
for bosons interacting via spherically-symmetric forces.
Dipolar length $a_{sd}$ modifies the isotropic  term and
it  governs the derivative term. By contrast to the so far employed
YY approximation, Eq.(\ref{Eq:DGPE}) is more concise
and remains valid even in a vicinity of scattering resonances.



To conclude, here I developed a new framework
for analyzing dipolar BECs. The principal results
are: the dipolar pseudopotential~(\ref{Eq:PS}), the energy functional~(\ref{Eq:EtotPsi}),
and the mean-field equation~(\ref{Eq:DGPE}). The results may be interpreted
in terms of the effective anisotropic mass: I showed that the
interactions between  dipoles
alter mass for a motion along the
polarizing field. For a typical BEC of spin-polarized magnetically-interacting
alkali-metal atoms
the effective mass is reduced by 10\% from it's bare value.
For a BEC of polarized heteronuclear molecules the mass may be reduced
by a factor of a 1,000.

This work was supported in part by the National Science Foundation and by the 
National Aeronautics and Space Administration under
Grant/Cooperative Agreement No. NNX07AT65A issued by the Nevada NASA
EPSCoR program.


\end{document}